\documentclass[lettersize, journal]{IEEEtran}
\usepackage[latin9]{inputenc}
\usepackage{color}
\usepackage{mathrsfs}
\usepackage{amsmath}
\usepackage{graphicx}

\makeatletter
\IEEEoverridecommandlockouts
\usepackage{cite}
\usepackage{amsfonts}\usepackage{algorithmic}
\usepackage{textcomp}
\usepackage{xcolor}
\def\BibTeX{{\rm B\kern-.05em{\sc i\kern-.025em b}\kern-.08em
    T\kern-.1667em\lower.7ex\hbox{E}\kern-.125emX}}

\usepackage{caption}
\usepackage{cite}
\usepackage{multirow}
\renewcommand{\maketag@@@}[1]{\hbox{\m@th\normalsize\normalfont#1}}

\usepackage{etoolbox}
\usepackage{xstring}\usepackage{mfirstuc}\usepackage{textcase}\usepackage[acronym]{glossaries}
\newcommand{\newac}{\newacronym}
\newcommand{\ac}{\gls}

\makeglossaries
\newac{isac}{ISAC}{integrated sensing and communications}
\newac{an}{AN}{artificial noise}
\newac{rcs}{RCS}{radar cross section}
\newac{csi}{CSI}{channel state information}
\newac{csit}{CSIT}{channel state information at transmitter}
\newac{fp}{FP}{fractional planning}
\newac{bme}{BME}{beampattern matching error}
\newac{ula}{ULA}{uniform linear array}
\newac{upa}{UPA}{uniform planar array}
\newac{cu}{CU}{communication user}
\newac{uavs}{UAVs}{unmanned aerial vehicles}
\newac{sinr}{SINR}{signal-to-interference-plus-noise ratio}
\newac{sdr}{SDR}{semi-definite relaxation}
\newac{1d}{1D}{one-dimensional}
\newac{zf}{ZF}{zero-forcing}
\newac{aod}{AoD}{angle of departure}
\newac{los}{LoS}{line-of-sight}
\newac{nlos}{NLoS}{non-line-of-sight}
\newac{bs}{BS}{base station}
\newac{bss}{BSs}{base stations}
\newac{cus}{CUs}{communication users}
\newac{snr}{SNR}{signal-to-noise ratio}
\newac{evd}{EVD}{eigenvalue decomposition}
\newac{qsdp}{QSDP}{quadratic semidefinite programing}
\newac{sdp}{SDP}{semidefinite programing}
\newac{dof}{DoF}{degrees-of-freedom}
\newac{soc}{SOC}{second order cone}
\newac{qos}{QoS}{quality-of-service}
\newac{lmi}{LMI}{linear matrix inequality}
\newac{fdd}{FDD}{frequency division duplex}
\newac{cscg}{CSCG}{circularly symmetric complex Gaussian}
\newac{crb}{CRB}{Cram{\'e}r-Rao bound}
\newac{sca}{SCA}{successive convex approximation}

\makeatother

\begin{document}
\title{Sensing-Assisted Sparse Channel Recovery for Massive Antenna Systems}
\author{Zixiang Ren, \textit{Graduate Student Member, IEEE,} Ling Qiu, \textit{Member,
IEEE,} Jie Xu, \textit{Senior Member, IEEE,} and Derrick Wing Kwan
Ng, \textit{Fellow, IEEE}\thanks{Z. Ren is with Key Laboratory of Wireless-Optical Communications,
Chinese Academy of Sciences, School of Information Science and Technology,
University of Science and Technology of China, Hefei 230027, China,
and the Future Network of Intelligence Institute (FNii), The Chinese
University of Hong Kong (Shenzhen), Shenzhen 518172, China (e-mail:
rzx66@mail.ustc.edu.cn).}\thanks{L. Qiu is with Key Laboratory of Wireless-Optical Communications,
Chinese Academy of Sciences, School of Information Science and Technology,
University of Science and Technology of China, Hefei 230027, China
(e-mail: lqiu@ustc.edu.cn).}\thanks{J. Xu is with the School of Science and Engineering (SSE) and the
FNii, The Chinese University of Hong Kong (Shenzhen), Shenzhen 518172,
China (e-mail: xujie@cuhk.edu.cn).}\thanks{D. W. K. Ng is with the University of New South Wales, Sydney, NSW
2052, Australia (e-mail: w.k.ng@unsw.edu.au).}\thanks{L. Qiu and J. Xu are the corresponding authors.}\vspace{-0.6cm}}
\maketitle
\begin{abstract}
This correspondence presents a novel sensing-assisted sparse channel
recovery approach for massive antenna wireless communication systems.
We focus on a fundamental configuration with one massive-antenna base
station (BS) and one single-antenna communication user (CU). The wireless
channel exhibits sparsity and consists of multiple paths associated
with scatterers detectable via radar sensing. Under this setup, the
BS first sends downlink pilots to the CU and concurrently receives
the echo pilot signals for sensing the surrounding scatterers. Subsequently,
the CU sends feedback information on its received pilot signal to
the BS. Accordingly, the BS determines the sparse basis based on the
sensed scatterers and proceeds to recover the wireless channel, exploiting
the feedback information based on advanced compressive sensing (CS)
algorithms. Numerical results show that the proposed sensing-assisted
approach significantly increases the overall achievable rate than
the conventional design relying on a discrete Fourier transform (DFT)-based
sparse basis without sensing, thanks to the reduced training overhead
and enhanced recovery accuracy with limited feedback. 
\end{abstract}


\section{Introduction}

\IEEEPARstart{D}{eploying} massive antennas at base stations (BSs)
has attracted a lot of attention in beyond fifth-generation (B5G)
and sixth-generation (6G) wireless networks. Such massive antenna
systems provide significantly increased spatial multiplexing, beamforming,
and diversity gains, as well as channel hardening effects, thus enhancing
data-rate throughput, lowering transmission latency, and improving
communication reliability. To fully reap these benefits, it is imperative
for the massive-antenna BS to acquire accurate channel state information
(CSI). This, however, presents practical challenges, especially for
downlink systems. For instance, conventional massive antenna systems
employ pilot-based channel estimation relying on the minimum mean
squared error (MMSE) principle, which induces significant pilot overheads
corresponding to the substantial quantity of transmit antennas \cite{minn2006optimal}.
To overcome this challenge, various prior works (see, e.g., \cite{lu2014overview})
have advocated reducing the pilot overheads and enhancing the communication
performance by utilizing the inherent sparsity of massive antenna
channels resulting from the limited scatterers in the environment,
especially in high frequency bands such as millimeter wave (mmWave)
and terahertz (THz).

Sparse channel estimation is implemented based on compressed sensing
(CS) techniques \cite{rao2014distributed,gao2015spatially,ding2018dictionary,zhang2023near}.
In this paradigm, the BS first transmits a limited number of pilots
(i.e., fewer than the large number of antennas). Subsequently, after
receiving the pilot signals, the CU sends back processed pilot information
to the BS. By capitalizing on the sparse nature of massive-antenna
channels and based on the limited feedback, the BS can recover the
wireless channel via well-established CS algorithms. For instance,
the authors in \cite{rao2014distributed,gao2015spatially} presented
basic pursuit (BP) based CS methods for sparse channel estimation,
in which the discrete Fourier transform (DFT) matrix is exploited
as the sparse basis for representing the channel. Furthermore, the
authors in \cite{ding2018dictionary} proposed a dictionary learning
approach to dynamically select a sparse basis from an overcomplete
DFT matrix. Nonetheless, this method suffers from the high computational
complexity of the overcomplete DFT matrix and the associated overhead
of dictionary learning. In addition, recent work \cite{zhang2023near}
studied the representation and estimation of sparse channels in the
near-field by considering the sparsity in both distance and angular
domains. However, these prior designs may suffer from compromised
performance and/or enhanced computational complexity due to the heuristically
chosen sparse basis (e.g., the over-complete DFT matrix) and the additional
cost of dictionary learning. Therefore, selecting an appropriate sparse
basis for concise sparse channel representation remains an essential
yet challenging task.

Recently, \ac{isac} has emerged as a crucial technology for 6G
wireless networks, where radar sensing is integrated into wireless
communications to enhance resource utilization efficiency and foster
mutual benefits \cite{liu2022integrated}. Among various ISAC design
paradigms, exploiting environmental sensing to assist channel estimation
and wireless communications is particularly appealing. For example,
the authors in \cite{liu2020joint} proposed a strategy where the
BS sends downlink pilots and conducts target sensing, while the CU
transmits uplink pilots. This strategy enables the BS to estimate
the downlink communication channel by jointly exploiting downlink
sensing results and received uplink pilots. Meanwhile, the authors
in \cite{li2022downlink} explored a scenario involving practical
codebook feedback. Here, the BS transmits downlink pilots and performs
target sensing, while the CU estimates the downlink channel and subsequently
provides practical codebook feedback to the BS. Furthermore, the authors
in \cite{huang2022joint} jointly investigated the target detection
and channel estimation problem via the common sparsity of communication
and sensing scatterers by jointly utilizing both uplink and downlink
pilots. Nevertheless, \cite{liu2020joint,li2022downlink,huang2022joint}
share a common challenge that the BS needs to transmit a substantial
number of pilots (exceeding the antenna count). By combining radar
sensing and sparse channel estimation, we can precisely identify a
proper sparse basis for CS signal recovery, thus motivating our work.

This correspondence proposes leveraging ISAC for efficient sparse
channel estimation in massive antenna systems with radar sensing.
We focus on a fundamental configuration featuring one massive-antenna
BS and one single-antenna CU. Within this framework, the BS simultaneously
transmits downlink pilots to the CU while receiving echo signals for
scatterer sensing, and then the CU provides feedback on its received
pilots to the BS. Leveraging this feedback, the BS identifies a sparse
basis and employs CS algorithms to accomplish channel recovery. Our
numerical results confirm the superiority of the sensing-assisted
approach over conventional designs relying on a DFT-based sparse basis
without sensing in terms of the overall achievable rate, thanks to
the reduced training overhead and improved accuracy with limited feedback.

\textit{Notations}: We use boldface lower- and upper-case letters
to denote vectors and matrices, respectively. The space of $N\times M$
complex matrices is represented by $\mathbb{C}^{N\times M}$. $\boldsymbol{I}$
stands for an identity matrix, while $\boldsymbol{0}$ represents
an all-zero matrix with appropriate dimensions. For a complex arbitrary-size
matrix $\boldsymbol{B}$, we use $\textrm{rank}(\boldsymbol{B})$,
$\boldsymbol{B}^{T}$, $\boldsymbol{B}^{H}$, and $\boldsymbol{B}^{c}$
to denote its rank, transpose, conjugate transpose, and complex conjugate,
respectively. $\mathcal{CN}(\boldsymbol{x},\boldsymbol{Y})$ denotes
a \ac{cscg} random vector with mean vector $\boldsymbol{x}$ and
covariance matrix $\boldsymbol{Y}$. The Euclidean norm of a vector
is represented by $\|\cdot\|$. $\|\cdot\|_{0}$ denotes the zero-norm
of a vector. $\mathcal{U}(\cdot)$ denotes a uniformly distributed
random variable. $\boldsymbol{A}\otimes\boldsymbol{B}$ represents
the Kronecker product of two matrices $\boldsymbol{A}$ and $\boldsymbol{B}$.
$\mathrm{diag}(\cdot)$ denotes a diagonal matrix with all non-diagonal
elements being zeros, and the diagonal elements determined by the
input. 

\section{System Model }

\begin{figure}
\centering\includegraphics[scale=0.26]{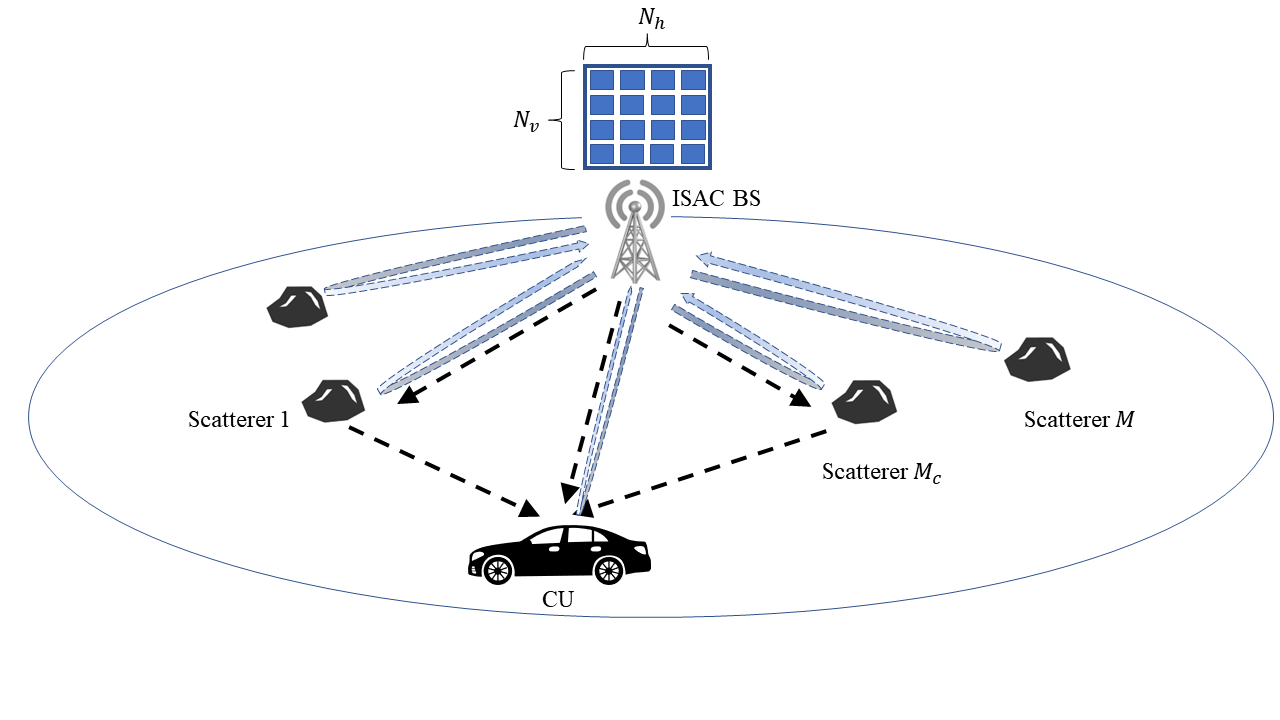}\centering\caption{\label{fig:1}Illustration of the massive antenna system.}
\vspace{-0.5cm}
\end{figure}

Fig. \ref{fig:1} shows a sensing-assisted massive antenna communication
system that comprises a multi-antenna \ac{isac} BS featuring a
\ac{upa} of $N_{v}\times N_{h}$ transmit antennas communicating
with a single-antenna CU\footnote{Extending this approach to multi-user or multi-antenna CUs remains
an area for future exploration.}. Here, $N_{v}$ and $N_{h}$ denote the vertical and horizontal antenna
numbers, respectively. Within the wireless environment, there are
$M$ scatterers, denoted by set $\mathcal{M}=\{1,\dots,M\}$. It is
assumed that only a subset of the environmental scatterers, identified
by set $\mathcal{M}_{c}=\{1,\cdots,M_{c}\}\subseteq\mathcal{M}$,
render a significant impact on the communication channel, while other
paths are blocked or ignored, in line with earlier studies \cite{liu2020joint,li2022downlink}.
As a result, the channel from the BS to the CU is expressed as\cite{li2022downlink}\vspace{-0.1cm}
\begin{equation}
\boldsymbol{h}=\sum_{m=1}^{M_{c}}\alpha_{m}\boldsymbol{a}(\theta_{m},\varphi_{m})=\sum_{m=1}^{M}\alpha_{m}\boldsymbol{a}(\theta_{m},\varphi_{m}),
\end{equation}
where $\alpha_{m}\in\mathbb{C}$ denotes the channel coefficient associated
with scatterer $m$, incorporating the signal propagation path loss
and the scatterer's \ac{rcs}, with\vspace{-0.1cm}
\begin{equation}
\begin{cases}
\alpha_{m}\neq0, & m\in\mathcal{M}_{c},\\
\alpha_{m}=0, & m\in\mathcal{M},m\notin\mathcal{M}_{c}.
\end{cases}
\end{equation}
Here, $\theta_{m}$ and $\varphi_{m}$ denote the associated elevation
and azimuth angles of departure of path $m$, respectively, and $\boldsymbol{a}(\cdot)$
denotes the steering vector of the transmit antenna array, i.e., \vspace{-0.08cm}
\begin{equation}
\begin{aligned} & \boldsymbol{a}_{v}(\theta_{m})=\frac{1}{N_{v}}[1,e^{j2\pi\frac{d_{v}}{\lambda}\sin\theta_{m}},\dots,e^{j2\pi\frac{d_{v}}{\lambda}(N_{v}-1)\sin\theta_{m}}]^{T},\\
 & \boldsymbol{a}_{h}(\theta_{m},\varphi_{m})\\
 & =\hspace{-0.1cm}\frac{1}{N_{h}}[1,e^{j2\pi\frac{d_{h}}{\lambda}\cos\theta_{m}\sin\varphi_{m}}\hspace{-0.1cm},\dots,e^{j2\pi\frac{d_{h}}{\lambda}(N_{h}-1)\cos\theta_{m}\sin\varphi_{m}}]^{T},\\
 & \boldsymbol{a}(\theta_{m},\varphi_{m})=\boldsymbol{a}_{v}(\theta_{m})\otimes\boldsymbol{a}_{h}(\theta_{m},\varphi_{m}).
\end{aligned}
\end{equation}
$\boldsymbol{a}_{v}(\theta_{m})$ and $\boldsymbol{a}_{h}(\theta_{m},\varphi_{m})$
represent the steering vectors related to the elevation and azimuth
angular perturbations, respectively, where $\lambda$ represents the
wavelength, while $d_{v}$ and $d_{h}$ represent the spacing between
two vertically and horizontally adjacent antennas, respectively. The
number of scatterers influencing the communication channel is often
limited due to the restricted angle spread \cite{sohrabi2021deep,gao2015spatially}.
Regarding this characteristic, researchers have advocated the exploration
of sparsity in the angular domain to reduce the training overhead
\cite{gao2015spatially}. In this context, CS is often regarded as
a promising method.

\subsection{Conventional CS-Based Sparse Channel Estimation}

In the conventional approach, the BS first transmits downlink pilots
to the CU. Subsequently, the CU provides feedback on the received
pilots\footnote{We consider the frequency division duplex (FDD) systems, in which
the conventional channel reciprocity is generally not applicable.}. The BS then proceeds to estimate the channel by exploiting the feedback
through CS. Finally, the BS transmits data based on the estimated
channel \cite{gao2015spatially}. Let us assume that the total coherent
block length is $T$ and the length of downlink pilots is $K$. The
total received downlink pilots by the CU are denoted as 
\begin{equation}
\boldsymbol{y}_{d}=\boldsymbol{X}_{d}\boldsymbol{h}+\boldsymbol{z}_{d},
\end{equation}
where $\boldsymbol{X}_{d}\in\mathbb{C}^{K\times N_{v}N_{h}}$ represents
the transmitted downlink pilots and $\boldsymbol{z}_{d}\in\mathbb{C}^{K\times1}$
is the Gaussian noise term, i.e., $\boldsymbol{z}_{d}\sim\mathcal{CN}(0,\sigma^{2})$
with $\sigma^{2}$ denoting the noise power. After receiving $\boldsymbol{y}_{d}$,
the CU feeds the quantized version $\bar{\boldsymbol{y}_{d}}$ back
to the BS. 

In order to recover the CSI based on $\bar{\boldsymbol{y}_{d}}$,
the BS exploits the sparsity with basis $\boldsymbol{A}_{d}=\boldsymbol{A}_{v}\otimes\boldsymbol{A}_{h}$,
where $\boldsymbol{A}_{v}$ and $\boldsymbol{A}_{h}$ are standard
discrete DFT matrices with dimensions $N_{v}$ and $N_{h}$, respectively.
Accordingly, the channel $\boldsymbol{h}$ is expressed as 
\begin{equation}
\boldsymbol{h}=\boldsymbol{A}_{d}\bar{\boldsymbol{\alpha}},
\end{equation}
where $\bar{\boldsymbol{\alpha}}\in\mathbb{C}^{N_{v}N_{h}\times1}$
is the sparse coefficients with sparse basis $\boldsymbol{A}_{d}$.
As a result, the conventional downlink channel estimation problem
utilizing CS is formulated as 
\begin{eqnarray}
\underset{\bar{\boldsymbol{\alpha}}\in\mathbb{C}^{N_{v}N_{h}\times1}}{\textrm{arg min}} & \|\bar{\boldsymbol{\alpha}}\|_{0} & \textrm{s.t. }\|\bar{\boldsymbol{y}_{d}}-\boldsymbol{X}_{d}\boldsymbol{A}_{d}\bar{\boldsymbol{\alpha}}\|\leq\varepsilon,\label{eq:p1}
\end{eqnarray}
where $\varepsilon$ denotes the recovery tolerance. It should be
noted that the CS signal recovery problem (\ref{eq:p1}) is generally
considered to be NP-hard. As such, various greedy-based algorithms
are available to tackle this challenge, including orthogonal matching
pursuit (OMP) and sparsity adaptive matching pursuit (SAMP) \cite{do2008sparsity}.
In this particular scenario where the exact sparsity level information,
denoted as $S$, is unavailable, the SAMP algorithm holds more appeal
\cite{do2008sparsity}. Specifically, the SAMP algorithm comprises
an inner loop and an outer loop. The sparsity is progressively expanded
stage by stage in the outer loop. Within the inner loop, the estimated
sparsity from the outer loop is utilized for the recovery of the signal
(the channel in our context). For a more comprehensive understanding
of the SAMP algorithm and its application, please refer to the detailed
explanation provided in \cite{do2008sparsity}. Let $\bar{\boldsymbol{\alpha}}^{*}$
denote the obtained solution to problem (\ref{eq:p1}). We then obtain
the recovered channel as
\begin{equation}
\bar{\boldsymbol{h}}=\boldsymbol{A}_{d}\bar{\boldsymbol{\alpha}}^{*}.
\end{equation}

Subsequently, we adopt the maximum ratio transmission for downlink
data transmission, where the transmit beamforming vector is set to
be $\frac{\sqrt{P}\bar{\boldsymbol{h}}}{\|\bar{\boldsymbol{h}}\|}$
with $P$ being the maximum transmit power. Consequently, the overall
achievable rate is calculated as
\begin{equation}
R=\frac{T-K}{T}\log_{2}\big(1+\frac{P|\bar{\boldsymbol{h}}^{H}\boldsymbol{h}|^{2}}{\|\bar{\boldsymbol{h}}\|^{2}}\big).\label{eq:7}
\end{equation}

It is important to note that the effectiveness of sparse channel recovery
is intricately connected to sparse basis $\boldsymbol{A}_{d}$. In
particular, this choice significantly affects the sparsity level of
$\boldsymbol{h}$, which directly impacts the overall recovery performance.
This thus motivates us to determine an effective sparse basis through
radar sensing. 

\begin{figure}
\vspace{-0.8cm}\centering\includegraphics[width=9cm,height=5cm]{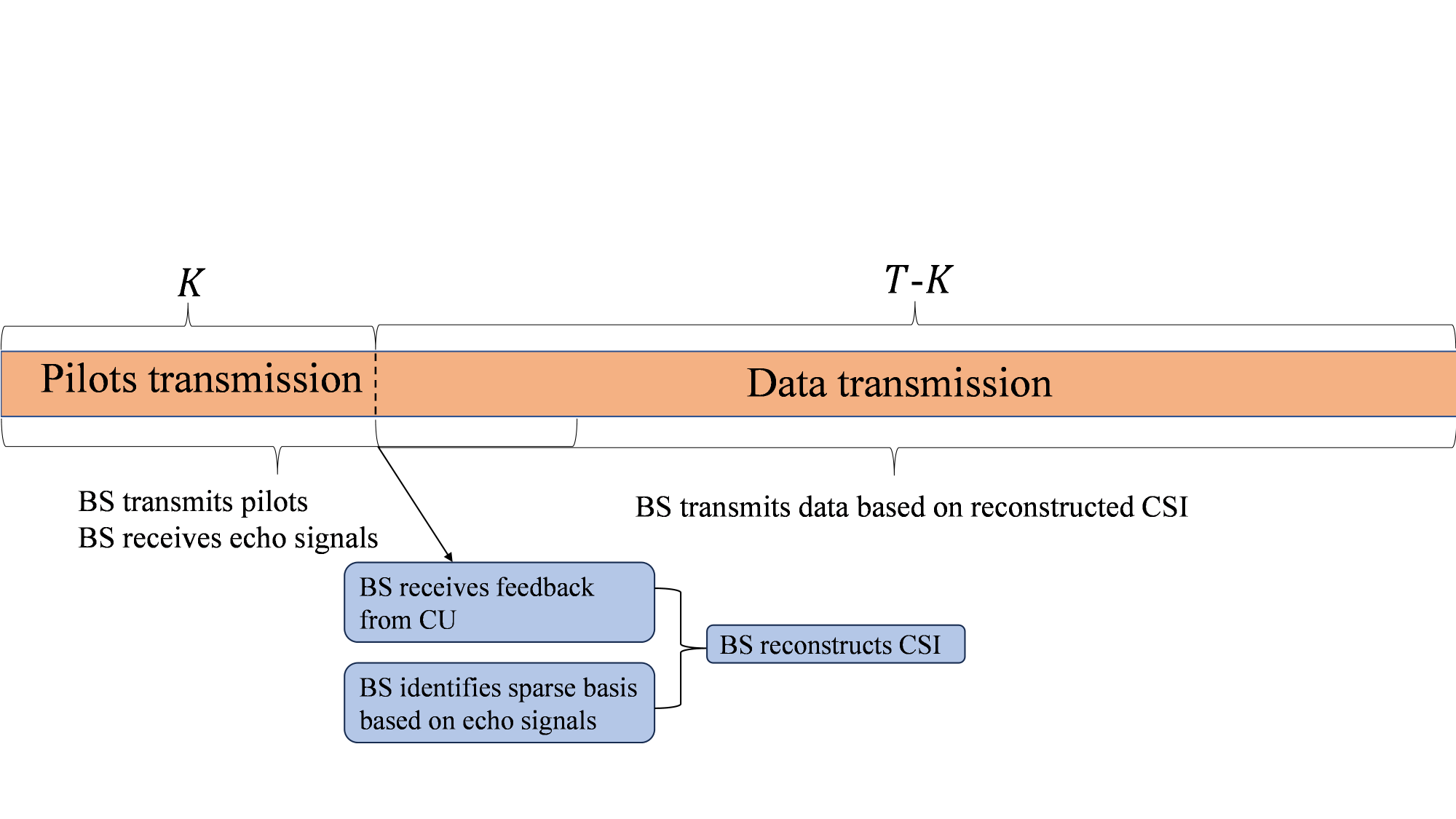}\caption{\label{fig:2}Transmission protocol for sensing-assisted sparse channel
estimation and wireless communications.}
\end{figure}

\section{Sensing-Assisted Sparse Channel Recovery }

This section proposes a sensing-assisted sparse channel recovery approach,
in which the BS accomplishes the sparse basis selection by acquiring
estimates of angles $\{\theta_{m}\}_{m=1}^{M}$, $\{\varphi_{m}\}_{m=1}^{M}$
via radar sensing. Subsequently, the BS reconstructs the CSI $\boldsymbol{h}$
by leveraging the complex coefficients $\{\alpha_{m}\}_{m=1}^{M}$
obtained through CS signal recovery. In this approach, we obtain environmental
side information through radar sensing. This enables us to dynamically
adjust the sparse basis, leading to performance improvements compared
to conventional designs.

In particular, we propose a framework for downlink transmission in
the massive antenna system assisted by radar sensing as shown in Fig.
\ref{fig:2}. The BS initially conducts light training pilot transmission
with a training length of $K$, and simultaneously receives reflected
echoes to estimate $\{\theta_{m}\}_{m=1}^{M}$, $\{\varphi_{m}\}_{m=1}^{M}$.
Next, the BS receives the channel feedback from the CU\footnote{Notably, considering the utilization of uplink resources for CU's
feedback, the time duration of feedback is not represented in Fig.
2.}. Consequently, the BS can reconstruct the channel vector as $\tilde{\boldsymbol{h}}$.
The overall achievable rate can be similarly calculated as \eqref{eq:7}
by replacing $\bar{\boldsymbol{h}}$ as $\tilde{\boldsymbol{h}}$.
In the sequel, we focus on the sparse channel recovery assisted by
radar sensing.

\subsection{Reconstruction of Sparse Basis via Radar Sensing}

In this subsection, we consider the radar sensing for sparse basis
reconstruction. To begin with, we focus on the downlink training pilots
transmission. Let $\boldsymbol{X}_{p}=[\boldsymbol{x}_{p}(1),\boldsymbol{x}_{p}(2),\dots,\boldsymbol{x}_{p}(K)]$
denote the transmitted pilots signal, where $\boldsymbol{x}_{p}(t)\in\mathbb{C}^{N_{v}N_{h}\times1},\forall t\in[1,\dots,K]$.
First, our attention turns to the radar sensing, where the BS employs
a colocated UPA consisting of $N_{v}\times N_{h}$ antennas for receiving
the echos and estimating the directions of $M$ scatterers. As a result,
the received echo signals at the BS in symbol $t$ are given as
\begin{equation}
\begin{array}{cr}
 & \boldsymbol{y}(t)=\sum_{m=1}^{M}\beta_{m}\boldsymbol{a}(\theta_{m},\varphi_{m})\boldsymbol{a}^{T}(\theta_{m},\varphi_{m})\boldsymbol{x}_{p}(t)+\boldsymbol{\boldsymbol{z}}(t),\\
 & t\in\{1,\dots,K\},
\end{array}\label{eq:Received signal for sensing}
\end{equation}
where $\beta_{m}$ denotes the reflection coefficient of the echo
channel associated with scatterer $m$, $\boldsymbol{\boldsymbol{z}}(t)\in\mathbb{C}^{N_{v}N_{h}\times1}$
denotes the received Gaussian noise, i.e., $\boldsymbol{\boldsymbol{z}}(t)\sim\mathcal{CN}(\boldsymbol{0},\sigma_{s}^{2}\boldsymbol{I})$
with $\sigma_{s}^{2}$ denoting the noise power. Let $\boldsymbol{Y}=[\boldsymbol{y}(1),\dots,\boldsymbol{y}(K)]$
denote the total received echo signals. Consequently, the BS can efficiently
estimate $\{\theta_{m}\}_{m=1}^{M},\{\varphi_{m}\}_{m=1}^{M}$ based
on the received echoes $\boldsymbol{Y}$ via different spatial signal
classification algorithms, such as multiple signal classification
(MUSIC) and estimation of signal parameters via rotational invariance
techniques (ESPRIT). Let $\{\hat{\theta}_{m}\}_{m=1}^{M}$ and $\{\hat{\varphi}_{m}\}_{m=1}^{M}$
denote the estimates of $\{\theta_{m}\}_{m=1}^{M}$ and $\{\varphi_{m}\}_{m=1}^{M}$,
respectively. Here, when the number of scatterers is significantly
smaller than that of transmit antennas, i.e., $M\ll N_{v}\times N_{h}$,
wireless channels exhibit sparsity in the angular domain \cite{gao2015spatially}.

Next, we identify the sparse basis from the estimated angles $\{\hat{\theta}_{m}\}_{m=1}^{M}$
and $\{\hat{\varphi}_{m}\}_{m=1}^{M}$. Let $\hat{\boldsymbol{A}}=[\boldsymbol{a}(\hat{\theta}_{1},\hat{\varphi}_{1}),\boldsymbol{a}(\hat{\theta}_{2},\hat{\varphi}_{2}),\dots,\boldsymbol{a}(\hat{\theta}_{M},\hat{\varphi}_{M})]$
and $J=\mathrm{rank}(\hat{\boldsymbol{A}})\leq M$. Suppose that the
singular value decomposition (SVD) of $\hat{\boldsymbol{A}}$ is given
by 
\begin{equation}
\hat{\boldsymbol{A}}=\boldsymbol{U}\boldsymbol{\varSigma}\boldsymbol{V}^{H},
\end{equation}
where $\boldsymbol{U}\in\mathbb{C}^{N_{v}N_{h}\times N_{v}N_{h}}$
and $\boldsymbol{V}\in\mathbb{C}^{M\times M}$ are unitary matrices,
and $\boldsymbol{\varSigma}=\left[\begin{array}{cc}
\boldsymbol{\varSigma}_{1} & \boldsymbol{0}\\
\boldsymbol{0} & \boldsymbol{0}
\end{array}\right]\in\mathbb{C}^{N_{v}N_{h}\times M}$ with $\boldsymbol{\varSigma}_{1}=\mathrm{diag}(\lambda_{1},\dots,\lambda_{J})$
and $\lambda_{1}\geq\dots\geq\lambda_{J}>0$. Moreover, we represent
the estimated channel $\tilde{\boldsymbol{h}}$ sparsely using an
orthogonal basis as 
\begin{equation}
\tilde{\boldsymbol{h}}=\hat{\boldsymbol{A}}\boldsymbol{\alpha}=\boldsymbol{U}\boldsymbol{\varSigma}\boldsymbol{V}^{H}\boldsymbol{\alpha}=\boldsymbol{U}\tilde{\boldsymbol{\alpha}},\label{eq:8}
\end{equation}
where $\boldsymbol{\alpha}\in\mathbb{C}^{M\times1}$ is the original
path scattering coefficients vector and $\tilde{\boldsymbol{\alpha}}=\boldsymbol{\varSigma}\boldsymbol{V}^{H}\boldsymbol{\alpha}$
represents the sparse vector to be recovered. Recall that $\boldsymbol{\varSigma}=\left[\begin{array}{cc}
\boldsymbol{\varSigma}_{1} & \boldsymbol{0}\\
\boldsymbol{0} & \boldsymbol{0}
\end{array}\right]\in\mathbb{C}^{N_{v}N_{h}\times M}$ and $\boldsymbol{V}^{H}\boldsymbol{\alpha}$ is an $M\times1$ vector,
there can be a maximum of $J$ non-zero elements within $\tilde{\boldsymbol{\alpha}}$.
As a result, we can adopt $\boldsymbol{U}$ as the sparse basis for
sparse signal recovery of $\tilde{\boldsymbol{\alpha}}$.

\subsection{Sparse Channel Recovery based on Feedback }

Then, the received signal at the CU is expressed as 
\begin{equation}
\boldsymbol{y}_{p}=\boldsymbol{X}_{p}\boldsymbol{h}+\boldsymbol{z},\label{eq:pilots}
\end{equation}
where $\boldsymbol{z}\in\mathbb{C}^{K\times1}$ denotes the Gaussian
noise at the CU receiver, i.e., $\boldsymbol{\boldsymbol{z}}\sim\mathcal{CN}(\boldsymbol{0},\sigma_{c}^{2}\boldsymbol{I})$,
where $\sigma_{c}^{2}$ is the noise power. The CU needs to extract
essential information from the received signal $\boldsymbol{y}_{p}$
and feed it back to the BS. Supposing that the CU feeds back $B$
bits of information, the feedback signal is expressed as 
\begin{equation}
\boldsymbol{q}=\mathscr{F}(\boldsymbol{y}_{p}),
\end{equation}
where function $\mathscr{F}(\cdot):\mathbb{C}^{K\times1}\rightarrow\{\pm1\}^{B}$
represents the adopted feedback scheme \cite{sohrabi2021deep}. In
particular, in this work, we employ a random vector quantization (RVQ)
codebook for the feedback of the received vector signal $\boldsymbol{y}_{p}$.
In this scheme, the CU first normalizes the vector $\boldsymbol{y}_{p}$
as $\bar{\boldsymbol{y}_{p}}=\frac{\boldsymbol{y}_{p}}{\|\boldsymbol{y}_{p}\|}$,
and then feeds back the codeword $\hat{b}$ satisfying 
\begin{equation}
\hat{b}=\underset{b\in\{1,2,\cdot\cdot\cdot,2^{B}\}}{\textrm{arg max}}|\bar{\boldsymbol{y}_{p}}^{H}\boldsymbol{c}_{b}|^{2},
\end{equation}
where $\boldsymbol{C}=[\boldsymbol{c}_{1},\boldsymbol{c}_{2},\dots,\boldsymbol{c}_{2^{B}}]\in\mathbb{C}^{K\times2^{B}}$
is the pre-defined $B$-bit RVQ codebook\footnote{RVQ has been widely adopted due to its ease of codebook construction
and suitability for low-rate feedback \cite{4100151}. It is worth
noting that other codebook methods, such as Grassmannian Manifolds
or DFT-based approaches, are also applicable.}. Assume that the BS can perfectly obtain the codeword feedback $\hat{b}$
and $\hat{\boldsymbol{y}_{p}}$ denotes the vector in the codebook
mapped by the codeword $\hat{b}$. 

Based on the feedback $\hat{\boldsymbol{y}_{p}}$ together with the
sparse basis constructed via radar sensing in \eqref{eq:8}, we formulate
the CS signal recovery problem as 
\begin{equation}
\arg\underset{\tilde{\boldsymbol{\alpha}}}{\min}\|\tilde{\boldsymbol{\alpha}}\|_{0},\mathrm{\textrm{ s.t. }}\|\hat{\boldsymbol{y}_{p}}-\boldsymbol{X}_{p}\boldsymbol{U}\tilde{\boldsymbol{\alpha}}\|\leq\varepsilon.\label{eq:14}
\end{equation}
 By incorporating the sparsity basis $\boldsymbol{U}$, received feedback
$\hat{\boldsymbol{y}_{p}}$, and via applying the CS-based SAMP algorithm,
we can achieve accurate and effective reconstruction of the sparse
signal $\tilde{\boldsymbol{\alpha}}$. Let $\tilde{\boldsymbol{\alpha}}^{*}$
denote the obtained solution to problem \eqref{eq:14}. Consequently,
the channel vector $\tilde{\boldsymbol{h}}$ is efficiently constructed
via \eqref{eq:8} as 
\begin{equation}
\tilde{\boldsymbol{h}}=\boldsymbol{U}\tilde{\boldsymbol{\alpha}}^{*}.
\end{equation}

\section{Numerical Results}

In this section, we illustrate the performance of our proposed sensing-assisted
CSI recovery algorithm. We evaluate the effectiveness of our proposed
sensing-assisted recovery method by comparing it with the conventional
benchmark that relies on a DFT-based sparse basis \cite{do2008sparsity}.
We assume that the BS transmits at a constant power level and the
pilot length is set equally in both the proposed design and the benchmark
for a fair comparison. For both the benchmark and our proposed sparse
basis selection designs, we consider two scenarios with finite feedback
and perfect feedback, respectively. 
\begin{itemize}
\item \textbf{Finite feedback}: The CU feeds back the received signal $\boldsymbol{y}_{p}$
with a finite number of bits.
\item \textbf{Perfect feedback}: The CU feeds back the received signal $\boldsymbol{y}_{p}$
with an infinite number of bits, i.e., the feedback of $\boldsymbol{y}_{p}$
is perfect.
\end{itemize}
We evaluate the performance of our proposed sensing-assisted sparse
basis selection design with finite feedback and compare it with the
following schemes:
\begin{itemize}
\item Benchmark with finite feedback
\item Benchmark with perfect feedback
\item Upper bound with perfect CSI
\item Proposed design with perfect feedback.
\end{itemize}
In this context, we examine a massive antenna system where a BS is
equipped with a half-wavelength UPA antenna configuration with $N_{v}=N_{h}=8$.
The BS is located at $[0\textrm{ m, }0\textrm{ m, }10\textrm{ m}]$
in an environment with $M=6$ paths, similar as \cite{sohrabi2021deep,li2022downlink},
among which $M_{c}=4$ scatterers contribute to the communication
channel. We assume that the small-scale complex path gain of each
path follows a standard Gaussian distribution, and the distance $d_{m}$
between scatterer $m$ and the BS is uniformly distributed in $[80\textrm{ m,}120\textrm{ m}]$.
We model $\theta_{m}$ and $\varphi_{m}$ as uniform distributed random
variables, i.e., $\theta_{m}\sim\mathcal{U}(-5{^\circ},+5{^\circ})$
and $\varphi_{m}\sim\mathcal{U}(-60{^\circ},+60{^\circ})$, $m\in\mathcal{M}$,
similar as \cite{sohrabi2021deep}. First, for the sensing model,
the complex sensing path coefficient $\beta_{m}$ is calculated by
$|\beta_{m}|=\sqrt{\rho_{0}d_{m}^{-2}\times\gamma_{m}d_{m}^{-2}}$,
where $\rho_{0}$ is the reference path loss at distance $1\textrm{ m}$
and is set as $-40\textrm{ dB}$, and $\gamma_{m}$ is a Gaussian
distributed reflection coefficient associated with the \ac{rcs}.
The phase of $\beta_{m}$ is randomly sampled from $[-\pi,\pi]$.
Then, as for the communication model, we assume that the scatterer
$m=1$ is the desired CU and the complex multipath gain $\alpha_{m}$
is calculated by $|\alpha_{m}|=\sqrt{\rho_{0}d_{m}^{-2}\times\delta_{m}r_{m}^{-2}}$,
where $r_{m}$ is the distance between scatterer $m$ and the CU,
while $\delta_{m}$ is a Gaussian distributed reflection coefficient.
We consider a coherent block consisting of $T=200$ symbols, and the
first $K=16\ll N_{v}N_{h}$ symbols are adopted for pilots transmission,
unless further specified. We perform 1000 random channel realizations
for each figure to evaluate the average performance.
\begin{figure}
\centering\begin{minipage}{0.48\columnwidth}

\centering\includegraphics[scale=0.32]{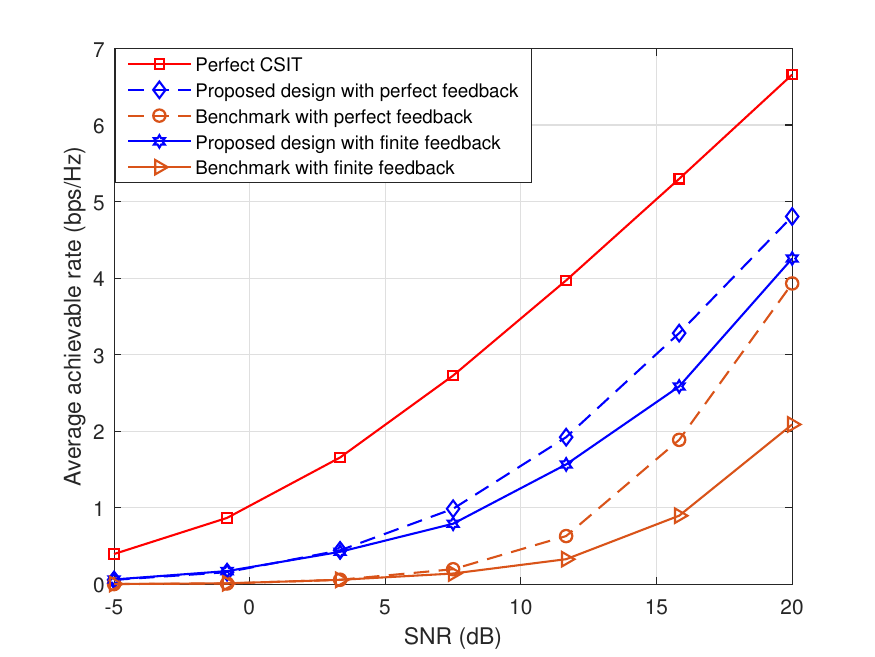}\caption{\label{fig:3}Average achievable rate versus different levels of receive
SNR, 12-bit RVQ feedback.}
\end{minipage}\hspace{+0.15cm}\centering\begin{minipage}{0.48\columnwidth}

\centering\includegraphics[scale=0.32]{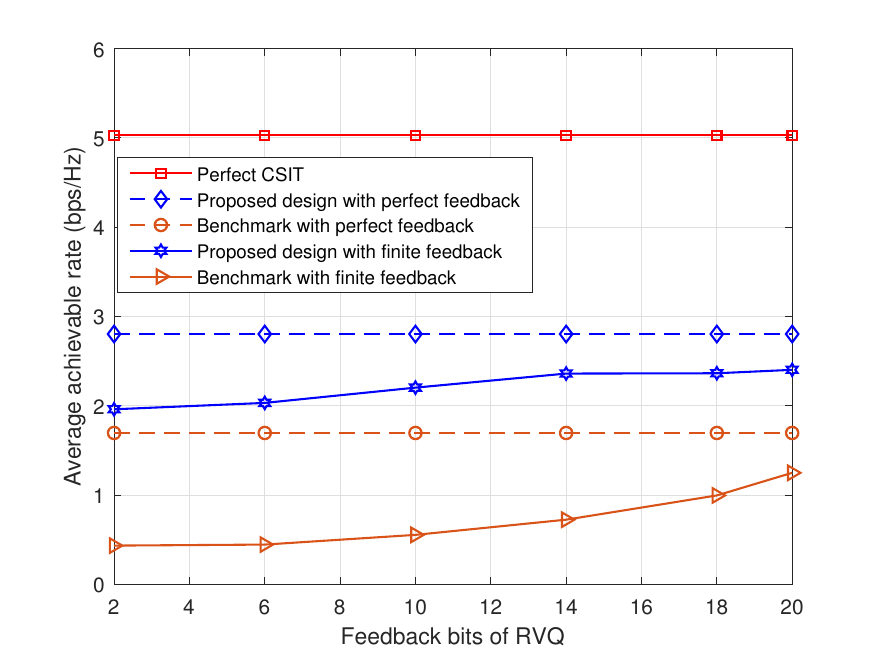}\caption{\label{fig:4}Achievable achievable rate versus different feedback
bits of RVQ, $\mathrm{SNR}=15\textrm{ dB}.$}
\end{minipage}\vspace{-0.2cm}
\end{figure}

Fig. \ref{fig:3} shows the average achievable rate versus the received
signal-to-noise ratios (SNR), represented as $10\log_{10}(\frac{P\|\boldsymbol{h}\|^{2}}{\sigma^{2}})$.
In the case of finite feedback, we consider the use of RVQ with 12
bits. It is observed that our proposed sensing-assisted sparse basis
consistently outperforms the conventional DFT based basis across all
three training and feedback scenarios. However, there is a performance
loss compared with the upper bound primarily due to the finite feedback
in both recovery algorithms. Furthermore, the finite feedback significantly
degrades the performance of the conventional DFT-based basis approach
due to the lower sparsity level of the feedback vector. 

Fig. \ref{fig:4} shows the achievable rate versus the different number
of feedback bits for RVQ. It is observed that our proposed sensing-assisted
scheme achieves satisfactory performance even with limited feedback
bits. This is attributed to the fact that we only need to recover
the signal within a small subspace, enabling a favorite sparsity level
and consequently leading to improved recovery performance. In contrast,
the conventional DFT based sparse basis exhibits poor sparsity, resulting
in a substantial decrease in recovery performance when utilizing finite
feedback. Consequently, the conventional scheme requires a larger
number of feedback bits for signal recovery compared to our proposed
sensing-assisted scheme.

\begin{figure}
\begin{minipage}{0.48\columnwidth}\centering

\centering\includegraphics[scale=0.32]{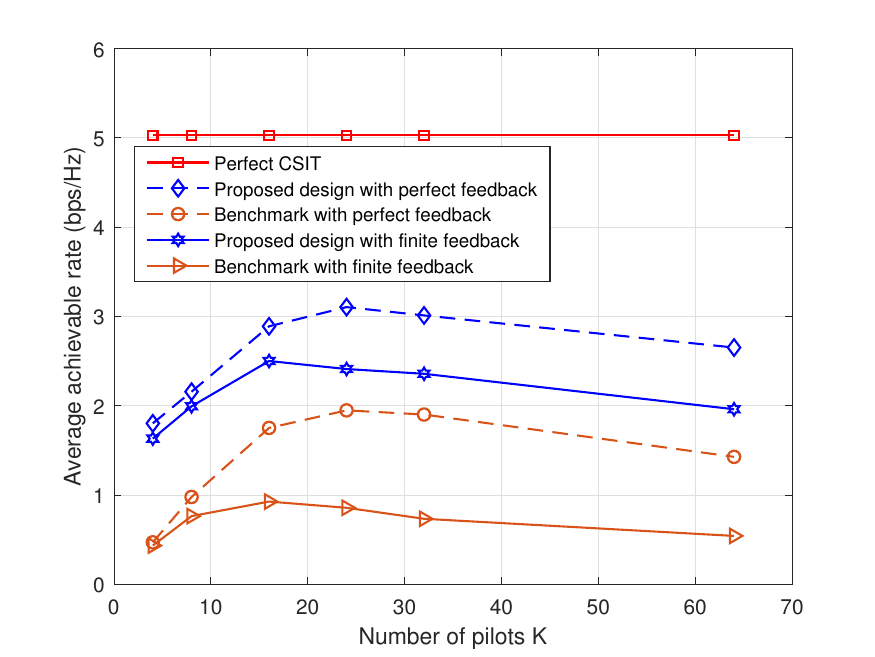}\caption{\label{fig:5}Average achievable rate versus different pilot lengths
$K$, $\mathrm{SNR}=15\textrm{ dB}$.}
\end{minipage}\hspace{+0.15cm}\begin{minipage}{0.48\columnwidth}\centering

\centering\includegraphics[scale=0.32]{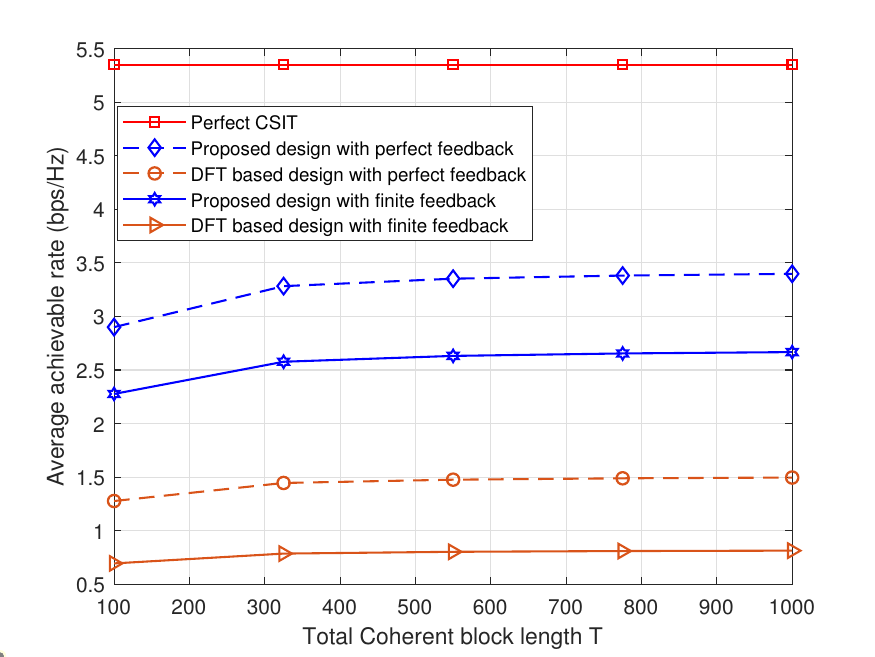}\centering\caption{\label{fig:6}Average achievable rate versus the total coherent block
length $T$, $\mathrm{SNR}=15\textrm{ dB}$.}
\end{minipage}\vspace{-0.2cm}
\end{figure}

Fig. \ref{fig:5} shows the average achievable rate versus the pilot
length $K$. It is observed that the achievable rate initially rises
and subsequently declines with an increasing pilot lengths $K$. This
is due to the fact that a higher number of pilots can lead to a more
accurate channel estimation, but can also reduce the block length
available for information transmission that outweighs the benefits.

Finally, Fig. 6 shows the average achievable\textcolor{blue}{{} }rate
versus the total coherent block length $T$. It is observed that the
achievable rate initially rises as $T$ increases and then becomes
saturated. This happens because the influence of the fixed pilot length
becomes negligible when $T$ is sufficiently large.

\section{Conclusion}

This correspondence presented an innovative approach for sparse channel
recovery in massive antenna wireless communication systems, leveraging
radar sensing. Our method integrated the transmission of downlink
pilots with scatterer sensing, user feedback reception, and the utilization
of echo sensing signals for CSI reconstruction via CS-based algorithms.
Numerical results highlighted substantial performance enhancements,
including a notable reduction in training overhead and a diminished
dependence on user feedback when compared to conventional methods
that solely rely on a DFT-based sparse basis. An interesting direction
for future research lies in extending the application of sensing-assisted
sparse channel recovery to distributed or multi-user scenarios, promising
to further enhance the versatility and efficacy of this approach.

{\footnotesize{}{} \bibliographystyle{IEEEtran}
\bibliography{IEEEabrv,IEEEexample,my_ref,IEEEfull}
 }{\footnotesize\par}

\end{document}